\documentclass[11pt]{article}
\usepackage{graphicx,amsmath}

\newcommand{\BABARProcNumber} {01/91}
\newcommand{\SLACPubNumber} {9064}

\input babarsym

\begin{document}
{\thispagestyle{empty}

\begin{flushright}
SLAC-PUB-\SLACPubNumber \\
\babar-PROC-\BABARProcNumber  \\
November, 2001 \\
\end{flushright}

\par\vskip 4cm

\begin{center}
\Large \bf {Results on mixing in the $D^0$ system from \babar}
\end{center}
\bigskip

\begin{center}
\large 
Monika Grothe\\
University of California, Santa Cruz, Santa Cruz CA 95064, USA \\
monika@scipp.ucsc.edu \\
(for the \lbabar\ Collaboration)
\end{center}
\bigskip \bigskip

\noindent
With 12.3~\invfb collected by the \babar\ experiment in 2001, the mixing 
parameter $y = \Delta \Gamma / 2 \Gamma$ is determined from the 
ratio of the $D^0$ lifetimes measured in the $D^0 \rightarrow K^- \pi^+$
and in the $D^0 \rightarrow K^- K^+$ decay modes. The preliminary result
$y = (-1.0 \pm 2.2({\rm stat.}) \pm 1.7({\rm syst.}))$ \% is obtained.
Also presented is the status of measuring the mixing parameters $y$ and 
$x^2 = (\Delta M / \Gamma)^2$ from a simultaneous fit to the time evolution
of the decay time distributions of Cabibbo-favored right-sign 
($D^0 \rightarrow K^- \pi^+$) and doubly Cabibbo-suppressed wrong-sign 
($D^0 \rightarrow K^+ \pi^-$) decays.
The wrong-sign decay rate, 
$R_{WS} =$ (\#  \ WS \  decays )  /  (  \# \ RS \ decays ) 
$= (0.38 \pm 0.04 ({\rm stat.}) \pm 0.02({\rm syst.}))$ \%  is obtained
from the fit to 23~\invfb of \babar\ data taken in 2000.

\vfill
\begin{center}
Contributed to the Proceedings of the \\
9$^{th}$ International Symposium 
On Heavy Flavor Physics \\
09/10/2001---09/13/2001, Pasadena, California, USA
\end{center}

\vspace{1.0cm}
\begin{center}
{\em Stanford Linear Accelerator Center, Stanford University, 
Stanford, CA 94309} \\ \vspace{0.1cm}\hrule\vspace{0.1cm}
Work supported in part by Department of Energy contract DE-AC03-76SF00515.
\end{center}

\section{Introduction}

If \CP conservation holds in the $D^0$ system, the \CP-even and \CP-odd
eigenstates are mass eigenstates with masses $M_+$, $M_-$ and widths 
$\Gamma_+$, $\Gamma_-$. The mixing parameters
$x = 2(M_+-M_-)/(\Gamma_++\Gamma_-)$ and 
$y = (\Gamma_+-\Gamma_-)/(\Gamma_++\Gamma_-)$ measure the difference 
of these masses and lifetimes, respectively. 
Since in the Standard Model $D$ - $\bar{D}$ mixing
is doubly Cabibbo-suppressed and vanishes in the $SU(3)$ flavor limit  
both parameters are expected to be small ($10^{-3})$~\cite{theory}.
If $D$ - $\bar{D}$ mixing were found to be large, this might be a sign either 
for large $SU(3)$ flavor breaking ($y \gsim x$) or for new physics 
($ x \gg y$)~\cite{Ligeti}.

The mixing parameter $y$ can be determined from the lifetime of
$D^0$ mesons\footnote{In the following, the charge conjugates of the 
$D^0$ meson and its decay modes are always implied as well.}
that decay into final states of specific \CP symmetry \cite{phenom}.
A final state that is an equal mixture of \CP-even and \CP-odd 
is produced by the Cabibbo-favored decay $D^0 \rightarrow K^- \pi^+$.
If $y$
is small, the lifetime distribution of $D^0$ mesons decaying into
this final state can be approximated as an exponential with
lifetime $\tau_{K\pi} = 1/\Gamma$ where $\Gamma = (\Gamma_+ + \Gamma_-)/2$.
A \CP-even final state is
produced by the singly Cabibbo-suppressed decay $D^0 \rightarrow K^- K^+$.
The decay time distribution of $D^0$ mesons that decay
into $K^- K^+$ is exponential
with a lifetime $\tau_{KK} = 1/\Gamma_+$. 
This lifetime can be compared
to $\tau_{K\pi}$ to obtain $y$:
\begin{equation}
y = \frac{\tau_{K\pi}}{\tau_{KK}} - 1 \;.
\label{eq:ycp}
\end{equation}

The mixing parameters $x^2$ and $y$ can be determined simultaneously from the 
time evolution of the wrong-sign (WS) decay $D^0 \rightarrow K^+ \pi^-$.
The WS decay rate has contributions from the doubly Cabibbo-suppressed 
decay, described by a pure exponential, and from $D$ - $\bar{D}$ mixing, 
described by the same exponential modified by a coefficient quadratic in the
decay time $t$. A third term arises from the interference of the two, where
the exponential has a coefficient linear in $t$. The time evolution of the
WS decay rate is thus      
described by~\cite{Nir}:
\begin{equation}
\Gamma(t) \ \propto \ \exp(-t) \ [ \ R \ + \ \sqrt{R} \ y^\prime \ t \ + \ 1/4 \
( \ x^{\prime 2} \ + y^{\prime 2} \ ) \ t^2 \ ]. 
\label{eq:mix}
\end{equation}
Here, $t$ is given in units of the lifetime of the $D^0$. 
The parameters $x^\prime$ 
and $y^\prime$ are related to the mixing parameters $x$, $y$ by a rotation:
\begin{equation}
x^\prime \ = \ x  \ \cos{\delta} \ + y \ \sin{\delta}, \quad \quad
y^\prime \ = \ y  \ \cos{\delta} \ - x \ \sin{\delta}.
\end{equation} 
The phase of the rotation is the strong phase between the doubly 
Cabibbo-suppressed contribution and the one from mixing.

In the following, two results are presented. Firstly,  
a preliminary measurement is described of $y$ from the difference of the
$D^0$ lifetime determined in the $K^- \pi^+$ and $K^- K^+$ decay channels.
Secondly, a preliminary measurement of the WS decay rate of the 
$D^0$ is discussed, extracted from a full fit to the time evolution of the 
combined right-sign and wrong-sign $D^0$ decay rates.

The measurements are based on data collected with the \babar\  detector at the
\pep2 asymmetric $e^+e^-$ collider. 
Data taken on and off the $\Upsilon(4S)$ resonance are used. Their 
center-of-mass is boosted along the beam axis with a nominal Lorentz boost 
of $\beta \gamma = 0.56$. 
The size of the interaction point (IP)
transverse to the beam direction is typically 6~$\mu$m in
the vertical direction and 120~$\mu$m in the horizontal direction.

Candidates for $D^0$ particles are identified through the decay 
$D^{*+} \rightarrow D^0 \pi^+$. Charged particles are detected and their 
momenta are measured
by a combination of a 40-layer drift chamber (DCH) and a five-layer, 
double-sided, silicon vertex tracker (SVT), both operating in a
1.5~T solenoidal magnetic field. A ring-imaging Cherenkov detector (DIRC) is 
used for charged particle identification.
A detailed description of the \babar\
detector is available in~\cite{babar}. 

\section{Measurement of $y$}

The result is obtained from a sample of 12.4\invfb
of 2001 \babar\ data that were reconstructed with the most advanced tracking alignment
parameters and reconstruction algorithms.

\subsection{Event Selection}

The widths $\Gamma_-$ and $\Gamma_+$ are determined by fitting the decay time
distributions of independent samples of 
$D^0 \to K^- \pi^+$ and $D^0 \to K^- K^+$ decays.
The $D^0$ candidates for each sample are identified
by means of the charged particles in their final state.
The decay $D^{*+} \rightarrow D^0 \pi^+$ and $K^\pm$ particle identification
are used to suppress backgrounds.

$D^0$ candidates are selected by searching for pairs of tracks of
opposite charge and combined
invariant mass near the $D^0$ mass $m_D$. Each track
is required to contain a minimum number of SVT and DCH hits in order
to ensure reconstruction quality. The two daughter
tracks of the $D^0$ candidate are fitted to
a common vertex. The $\chi^2$ probability of this vertex fit has to be better 
than 1\%.

Each $K^\pm$ candidate among the $D^0$ daughter tracks 
is required to pass a likelihood-based particle
identification algorithm. This algorithm is based on the measurement
of the Cherenkov angle by the DIRC for momenta
$p \gsim 0.6$\gevc and on the energy loss ($dE/dx$) measured by 
SVT and DCH for momenta $p \lsim 0.6$\gevc.
An average efficiency for $K^\pm$ identification of 
approximately 80\% is reached 
for tracks within the DIRC acceptance while the $\pi^\pm$ misidentification
probability amounts to about 2\%.

The decay $D^{*+} \rightarrow D^0 \pi^+$ can be distinguished by a
$\pi^+$ of low momentum, commonly referred to as the slow pion
($\pi_s$). To increase acceptance, $\pi_s$ candidate tracks do not have to
contain DCH hits. To improve momentum resolution, a vertex fit
is used to constrain each $\pi_s$
candidate track to pass through the intersection of the
$D^0$ trajectory and the IP. If the $\chi^2$ probability of this
vertex fit is less than 1\%, the $D^*$ candidate is discarded.

The $D^*$ candidates peak at a value of
$\delta m \approx 145.4$\mevcc, where $\delta m$ is the difference
in the reconstructed $D^*$ and $D^0$ masses. Backgrounds are reduced
by rejecting events with a value of $\delta m$ that deviates more than 
a given margin from the peak. The size of this margin corresponds to
approximately three standard deviations and
varies between 1 and 2.5\mevcc, depending on the quality of the $\pi_s$ track.

\begin{figure}[!t]
\begin{center}
\includegraphics[width=0.75\linewidth]{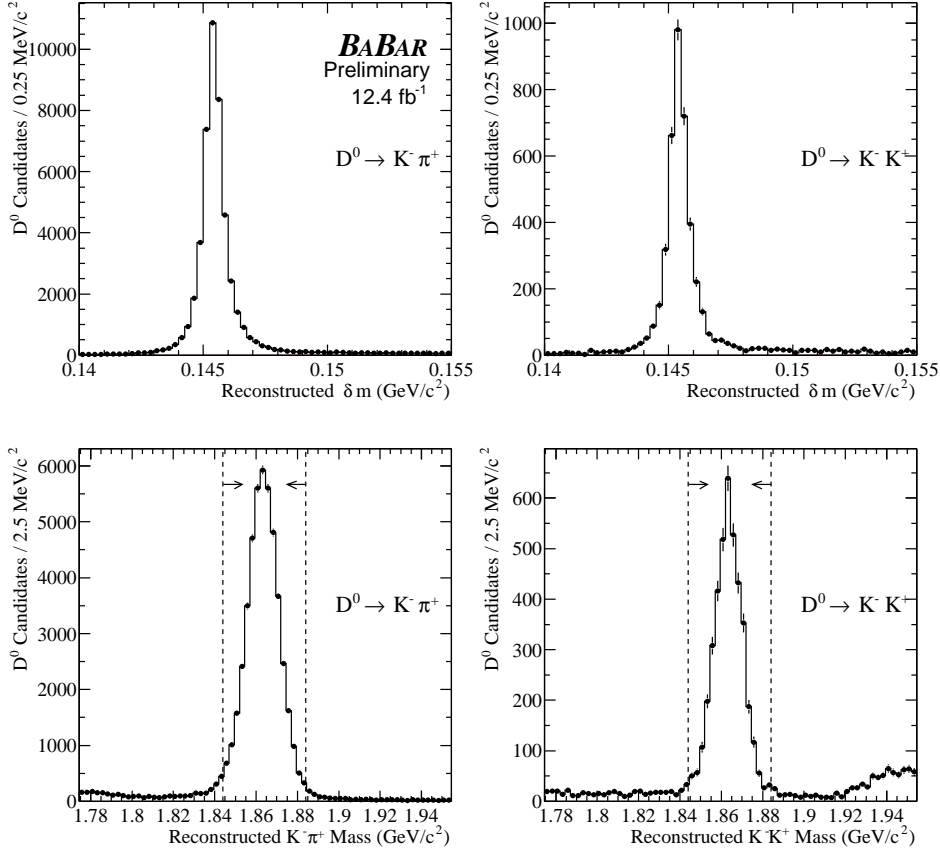}
\caption{
The reconstructed $\delta m$ and $D^0$ mass distributions
after event selection
for the $K^- \pi^+$ and $K^- K^+$ decay modes. The $\delta m$ plots
include candidates both inside and outside the $\delta m$ selection
requirement which fall within the $m_D$ window
indicated in the lower plots.
}
\label{fg:dist}
\end{center}
\end{figure}

To remove background from $B$ meson decays, each $D^*$ candidate
has to have a momentum $p^*$ greater than 2.6\gevc in the center-of-mass. 
This condition is also effective at removing
combinatorial background that tends to accumulate at lower momenta.

After all selection requirements, about 45,000 $D^0$ candidates in the
decay channel $K^- \pi^+$ and about 4,000 in the $K^- K^+$ decay channel are
retained. The mass and
$\delta m$ distribution of the selected events are shown in
Fig.~\ref{fg:dist}. The relative size of the remaining background is about
2\% and 5\% for the $K^-\pi^+$ and $K^-K^+$ samples, respectively,
when measured inside a $\pm 20$\mevcc mass window.

\subsection{Lifetime Determination}

The flight length and its measurement error are determined for each
$D^0$ candidate by a global,
three-dimensional,
multiple-vertex fit that includes the $D^0$ daughter tracks, 
the $\pi_s$ track,
and the IP envelope. This fit does not include explicit constraints
on the $D^*$ or $D^0$ masses. The value listed by the
Particle Data Group (PDG) for the $D^0$
mass ($m_D = 1.8654$~\gevcc \cite{pdg}) and the momentum of the $D^0$ obtained
with the vertex fit are used to calculate the boost of the $D^0$
and obtain the proper decay time.

The lifetimes of the $D^0$ candidate samples are extracted by means of an 
unbinned maximum likelihood fit.
The likelihood function is divided into two distinct
decay time distribution functions, one for the signal and one for the 
background. The signal
function is the convolution of an exponential and
a resolution function. The resolution function consists of the sum of
two Gaussian distributions with zero mean and with widths that are
proportional to the measurement error (typically 180~fs)
of the decay time of each $D^0$ candidate.
The parameters in the fit associated with the signal are
the lifetime, the proportionality factors for the widths of the two Gaussians,
and the fraction of signal that is assigned to the second Gaussian.

Like the signal likelihood function, 
the background function consists of the convolution of
a resolution function and a lifetime distribution.
The background lifetime distribution is determined as the sum of an
exponential distribution and a delta function at zero, the
latter corresponding to those sources of 
background that originate at the IP.
The resolution function consists of the sum
of three Gaussian distributions. The first two are
chosen to match the resolution function of
the signal. The third is given a
width independent of the decay time error and accounts for outliers
produced by long-lived particles or reconstruction errors.
The additional fit parameters associated with the background include the
fraction assigned to zero lifetime sources, 
the background lifetime, the width of the third
Gaussian, and the fraction of background assigned to the third
Gaussian.

The reconstructed mass of each $D^0$ candidate provides the likelihood 
of this candidate to be part of the signal.
This likelihood is based on a separate fit of the
reconstructed $D^0$ mass distribution.
This fit includes a resolution function composed of 
a Gaussian with an asymmetric tail
and a linear portion 
to describe the
background. The slope of the background part is constrained with
$D^0$ candidates in the $\delta m$ sideband ($151 < \delta m < 159$\mevcc).

\begin{figure}[!t]
\begin{center}
\includegraphics[width=0.75\linewidth]{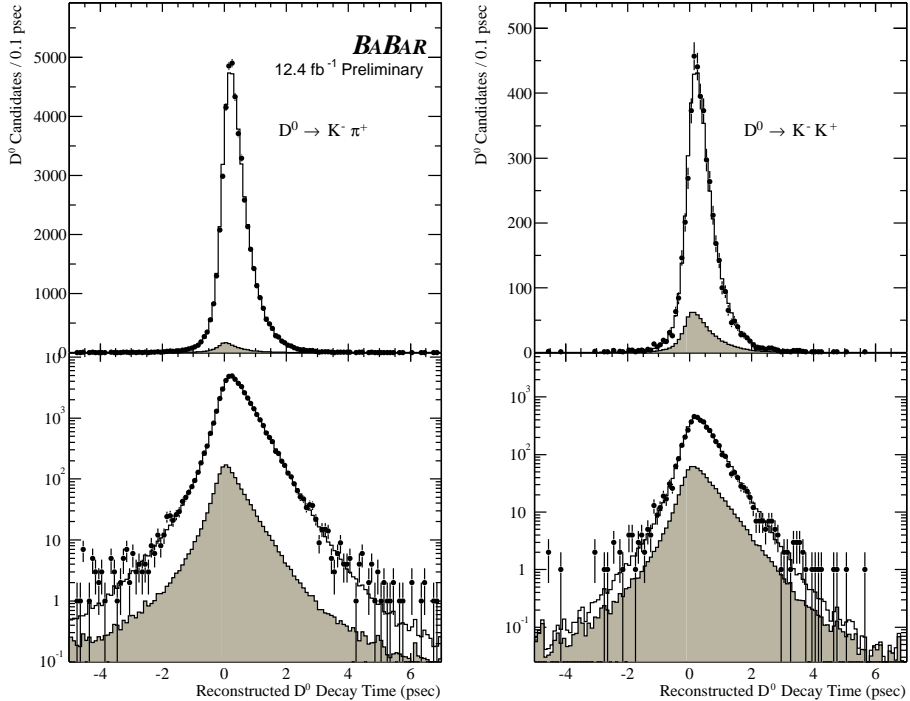}
\caption{
The fit to the reconstructed $D^0$ decay time distribution in the two 
$D^0$ decay modes and
for all events including the $D^0$ mass sidebands.
The white histogram represents the result of the unbinned maximum likelihood
fit described in the text. The gray histogram is the portion
assigned by the fit to background.
}
\label{fg:fit}
\end{center}
\end{figure}

The results of the lifetime fits are shown in Fig.~\ref{fg:fit}.
Typical values for the fitted parameters are
a background lifetime similar to the $D^0$ lifetime
and a width of the third Gaussian that is several times larger than
the typical decay time error. The proportionality factors associated with
the two Gaussians in the resolution function correspond to
a root-mean-square of approximately $1.2$.

To ensure that the analysis was performed in an objective manner,
the $D^0$ lifetime and $y$ values
were blinded throughout the analysis process. This blinding was
performed by adding to each of the $\tau_{K\pi}$ and $\tau_{KK}$ fit results
an offset chosen from a random Gaussian distribution
of width 10~fs. The values of the two offsets and
the positive ($\tau_i > 0$) sides of the lifetime
distribution of the data and of the fit result (Fig.~\ref{fg:fit})
were concealed. The value of $y$ was unblinded only
after the analysis method and systematic uncertainties
were finalized and the result was committed for public release.

\subsection{Systematic Errors and Result}

Many systematic uncertainties cancel
because $y$ is measured from the ratio of lifetimes.
The uncertainties that do not cancel
are associated mostly with backgrounds. These are tested by varying
each event selection requirement within its uncertainty and 
recording the subsequent change $\Delta_i$ in the measured value
of $y$.
The quadrature difference
$(\delta \Delta_i)^2 = | \sigma_0^2 - \sigma_i^2 |$
is used as an estimate of the
statistical error $\delta \Delta_i$ in $\Delta_i$,
where $\sigma_0$ ($\sigma_i$)
is the statistical error in $y$ before (after)
the $i{\hskip 0.5pt}$th systematic
check. Only systematic checks with $\Delta_i > \delta\Delta_i$
are included in the sum $\sum \Delta_i^2 - (\delta\Delta_i)^2$.
The square root of this sum (1.7\%) provides an estimate of the
systematic uncertainty originating from the event selection and
backgrounds.

Biases in tracking reconstruction
are investigated by
studying Monte Carlo samples, which, within statistics,
show no reconstruction bias. 
In addition, a variety of vertexing techniques, including constraining the
$D^0$ mass and using separate $D^*$ and $D^0$ vertex fits are employed as
systematic checks.
A systematic uncertainty of 0.4\% is assigned from this source.

Detector misalignment is another potential source of systematic uncertainty.
Residual internal misalignment of the SVT,
even as small as a few microns, can produce significant
variations in the apparent $D^0$ lifetime.

This source is studied with the help of
$e^+ e^- \rightarrow \gamma\gamma \rightarrow 4\pi^\pm$ events
in which the four charged tracks are known to originate from the IP.
By selecting oppositely charged pairs of these tracks with opening
angles similar to two-prong $D^0$ decays, it is possible to
measure the apparent IP position as a function of $D^0$ trajectory
and to calculate a correction to the $D^0$ lifetime obtained by the fit. 
For the data
samples used in this analysis, a correction of
$+5$~fs is found, with negligible statistical error and a systematic
uncertainty of $\pm5$~fs. This type of correction nearly
cancels in the lifetime ratio and introduces little systematic uncertainty
in $y$.

\begin{table}[!hpb]
\begin{center}
\begin{tabular}{lc}
\hline
\hline
Category & Uncertainty (\%)         \\
\hline
Event Selection and Background      & $1.7$ \\
Reconstruction and Vertexing        & $0.4$ \\
Alignment                           & $0.3$ \\
\hline
Quadrature Sum                      & $1.7$ \\
\hline
\hline
\end{tabular}
\end{center}
\caption{Individual contributions to the systematic uncertainty in $y$.}
\label{tab:syst}
\end{table}

The systematic uncertainties in $y$ 
are summarized in Table~\ref{tab:syst}.
Adding the contributions of all systematic checks in quadrature, 
the preliminary result for $y$ is obtained:
\begin{equation}
y = \left( -1.0 \pm 2.2 \pm 1.7 \right)\;\% \;,
\end{equation}
where the first error is statistical and the second systematic.
This result is consistent with the Standard Model expectation of zero
and with results from
E791 \cite{e791}, FOCUS \cite{focus}, Belle \cite{belle} 
and CLEO \cite{cleo}.

An important test of the soundness of the analysis is a 
$D^0$ lifetime that agrees with the PDG value of $412.6 \pm 2.8$~fs 
\cite{pdg}. A corrected value of 
$\tau_{K\pi} = 412 \pm 2$~fs is found. The systematic uncertainty is 
approximately 6~fs and is dominated by detector alignment effects.

\section{Measurement of the wrong-sign $D^0$ decay rate}

Wrong-sign (WS) decays of the type $D^0 \rightarrow K^+ \pi^-$ 
are selected by requiring that the slow pion and the kaon have the same charge.
The measurement is based on 23\invfb of BaBar data taken in 2000.
Vertex reconstruction, event selection and systematic error analysis 
are similar to the ones employed in the
measurement of $y$, described above. After all event selection requirements
a sample of about 200 WS candidate events is retained.  

\begin{figure}[!b]
\begin{center}
\includegraphics[width=0.5\linewidth]{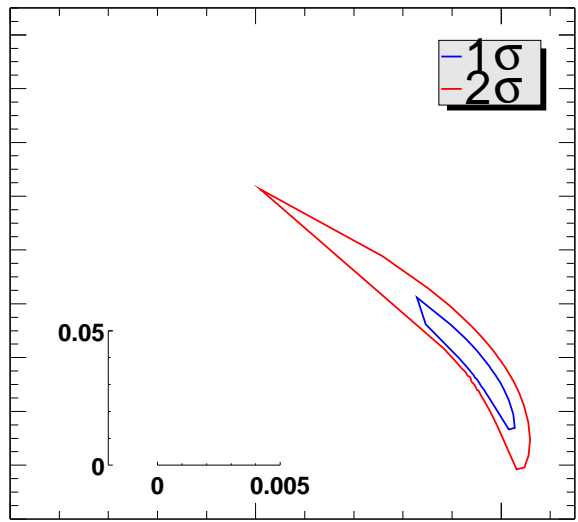}
\caption{The 1 and 2~$\sigma$ contours of the statistical error 
in the $x^{\prime 2}$-$y^\prime$-plane ($y^\prime$ on vertical axis) obtained from the simultaneous 
fit to the 
time evolution of RS and WS decay rates. A length scale is given for the axes but no axis labels and thus
no central values for $x^{\prime 2}$, $y^\prime$. 
}
\label{fg:x2y}
\end{center}
\end{figure}

A simultaneous unbinned maximum likelihood fit to the decay time distribution 
of right-sign (RS) and wrong-sign $D^0$ decay candidates is performed. 
The fit function consists of a lifetime distribution function convoluted with a
resolution function. 
For RS events, a pure exponential is assumed as functional form of 
their lifetime distribution while for WS events the functional form follows
Eq.~\ref{eq:mix}. 
The simultaneous fit of RS and WS candidates exploits the high statistics
of the RS sample to obtain an accurate description of the resolution function
and of the type and size of the background contributions.  

The 1 and 2~$\sigma$ contours of the statistical
error of the fit result in the $y^\prime$-$x^{\prime 2}$-plane (Fig.~\ref{fg:x2y}) 
exhibit a strong correlation
between the two parameters that extends into the unphysical region $x^{\prime 2} < 0$.
Note that a length scale is given for the axes but no axis labels and thus
no central values for $x^{\prime 2}$, $y^\prime$.
The analysis of systematic errors shows that the dominating systematic error
source in the 2000 \babar\ data is the internal alignment of the SVT.
It will soon be possible to extract the central values of $x^{\prime 2}$ and $y^\prime $
from more data
reconstructed with the most advanced tracking alignment constants.

The fit function contains as one parameter the WS decay rate, defined as:
\begin{center}
$R_{WS}$ = (number  of wrong-sign  decays) / (number of right-sign decays)
\end{center}
The simultaneous fit to the RS and WS candidate events yields the following 
result for the WS decay rate:
\begin{center}
$R_{WS}$ = ( 0.38 $\pm$ 0.04 $\pm$ 0.02) \%, 
\end{center}
where the first error is statistical, the second systematic. This result is 
compatible, within its errors, with the results obtained by E791~\cite{E791-2},
ALEPH~\cite{ALEPH-2}, CLEO~\cite{CLEO-2},
FOCUS~\cite{FOCUS-2}, Belle~\cite{belle} and is the  
most precise experimental measurement of $R_{WS}$ currently available.

\section{Conclusion}

With 12.4\invfb of data taken by the \babar\  detector
in 2001, the preliminary result for $y$ is obtained:
\begin{equation}
y = \left( -1.0 \pm 2.2 \pm 1.7 \right)\;\% \;,
\end{equation}
where the first error is statistical and the second systematic.
This result is consistent with the Standard Model expectation of zero
and is consistent with published values from
E791 \cite{e791} and FOCUS \cite{focus}
and preliminary results from Belle \cite{belle} and CLEO \cite{cleo}.
A reduction of the statistical error by half is feasible in the near future
by considering additional decay channels ($D^0 \rightarrow \pi^- \pi^+$)
and by including the 23~\invfb of \babar\ data collected in 2000.

The extraction of the mixing parameters $x^2$ and $y$ by measuring the 
time evolution of the wrong-sign ($D^0 \rightarrow K^+ \pi^-$) decay rate
in a sample of 35~\invfb of combined 2000 and 2001 \babar\ data will be 
possible soon. 

Already with the 23~\invfb of \babar\ data collected in 2000, the most precise
existing measurement of the wrong-sign decay rate 
$R_{WS} = ( \#  \ WS \  decays   )  /  (  \# \ RS \ decays  )$
is achieved:
\begin{equation} 
R_{WS} =  (0.38 \pm 0.04 \ ({\rm stat.}) \pm 0.02 \ ({\rm syst.})) \;\% \; .
\end{equation} 
This result is 
consistent with results by E791~\cite{E791-2},
ALEPH~\cite{ALEPH-2}, CLEO~\cite{CLEO-2},
FOCUS~\cite{FOCUS-2} and a preliminary result by Belle~\cite{belle}.

\end{document}